\documentstyle[tighten,pra,aps,epsfig,multicol]{revtex}

\begin{document}
\title{Optimal States for Bell inequality Violations using Quadrature 
Phase Homodyne Measurements}
\author{W.J. Munro}
\address{Centre for Laser Science,Department of Physics, University of
Queensland, QLD 4072,  Brisbane, Australia}
\date{\today}

\maketitle

\begin{abstract}
We identify what ideal correlated photon number states are to 
required to maximize the discrepancy between local realism and quantum mechanics 
when a quadrature homodyne phase measurement is used. Various Bell 
inequality tests are considered.
\end{abstract}

\begin{multicols}{2}
\section{Introduction}
There has been recently active interest in tests of quantum mechanics{\cite{EPR 35}} 
versus local realism in a high efficiency detection limit. 
Several authors{\cite{Gilchrist Deuar and Reid 98,Munro and Milburn 98,Yurke and Stoler 97}} including ourselves
have considered detection schemes quadrature phase homodyne 
measurements. Such schemes use strong local oscillators and hence have 
very high detection efficiency{\cite{Polzik Carry and Kimble 92}}. 
This removes one of the current loopholes{\cite{CHSH 69,Kwiat et al 94,Freyberger et ak 95,Fry et al 96}} 
and potentially allows a strong test of quantum mechanics{\cite{strong}} to be performed. 

The original idea of Gilchrist {\it et. al.}{\cite{Gilchrist Deuar and Reid 98}} was to use a 
{\it circle} or pair coherent state{\cite{Agarwal 86,Tara and Agarwal 
94,Reid and Krippner 93}} produced by 
nondegenerate parametric oscillation with the pump mode 
adiabatically eliminated. Using highly efficient quadrature 
phase homodyne measurements, the Clauser Horne strong Bell 
inequality{\cite{Bell 65,Clauser and Horne 74,Clauser and Shimony 78}} 
could be tested in an all optical regime. 
A small (approximately $1.5\%$) but significant theoretical violation was found 
for this extremely idea system.  While the mean photon number for the 
system may be low (approximately $1.12$), the use of homodyne 
measurements allow a macroscopic current to be detected.

In this article, we take an unphysical but interesting approach and answer 
the following questions: 

\begin{itemize}

\item {\it Given that your detection scheme is a quadrature phase homodyne 
measurement, what is the optimal input or correlated photon number 
state to maximize the potential violation?}

\item {\it What is the optimal Bell inequality to test?}
\end{itemize}
To begin we will restrict our attention to correlated photon number 
states of the form
\begin{eqnarray}\label{correlatedpair}
| \Psi \rangle = \sum_{n=0}^{\infty} c_{n} | n \rangle | n \rangle 
\end{eqnarray}
Two main sources of correlated photon number currently exist, 
each having it own particular form of $c_{n}$. The most well known is 
simply the nondegenerate parametric amplifier specified 
by an ideal  Hamiltonian of the form{\cite{Reid 89}}  
\begin{eqnarray}
H=-\hbar \chi \epsilon \left( a b+a^\dagger b^\dagger\right).
\end{eqnarray}
where $\epsilon$ is field amplitude of a nondepleting classical pump 
and  $\chi $ is proportional to the susceptibility of the medium.
$ a,b$ are the boson operators for the orthogonal signal 
and idler modes. After a time $\tau$, the state of the system is given 
by (\ref{correlatedpair}) with $c_{n}$ specified by
 \begin{eqnarray}
c_{n}={{\tanh^n \left[\chi \epsilon \tau\right]}\over{\cosh \left[\chi \epsilon \tau\right] }}
\end{eqnarray}
In the quadrature phase amplitude basis this state has a positive Wigner function. 
Hence it can be described as a local hidden variable theory and thus cannot violate a Bell 
inequality.

The other source of highly correlated photon number states exists in  nondegenerate 
parametric oscillation. In the limit of very large parametric 
nonlinearity and high Q cavities, a state of the form{\cite{Reid and Krippner 93}}
\begin{eqnarray}\label{circle}
| \Psi \rangle ={e^{r^{2}}\over \sqrt{4 \pi^2 I_{0}\left(2 r^{2} \right)}}
\int_{0}^{2 \pi} d \theta\; | r e^{i \theta}\rangle| r e^{-i \theta}\rangle 
\end{eqnarray}
can be generated. Here $r$ is the size of the circle of the coherent 
states and $I_{0}$ is the zeroth order modified Bessel function.  
Equivalently this state can be written in the form of (\ref{correlatedpair}) 
with $c_{n}$ given by
\begin{eqnarray}\label{circlecn}
c_{n}={{r^{2 n}}\over{n! I_{0}\left(2 r^{2} \right)}} 
\end{eqnarray}
This was the state considered by Gilchrist {\it et. al.}{\cite{Gilchrist Deuar and Reid 98}}

\begin{figure}[h]
\center{ \epsfig{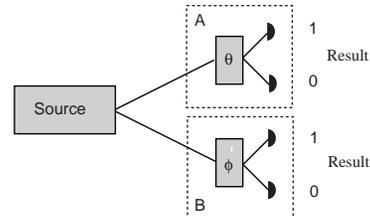}}
\caption{Schematic of a very generalized Bell experiment setup. After 
a source prepares two particles, these particles are directed out to 
the locations $A$ and $B$. At each location there is an analyzer with 
an adjustable parameters $\theta,\phi$. The particles are then 
detected resulting in a binary result ``1'' or ``0'' individually. These 
results can then be used to build up the statistics necessary to test the various Bell 
inequalities.}
\label{fig1}
\end{figure}

Given the general form of known correlated number states (\ref{correlatedpair}), the next fundamental 
question that should be initially addressed is what we 
mean by the Bell inequality. A number of Bell inequalities exist, and the particular 
one depends used heavy on your application and experimental setup. 
The Bell inequalities to be considered in this article are the Clauser 
Horne{\cite{Clauser and Horne 74}}, the {\it spin}{\cite{Bell 65}}, and the 
information-theoretic{\cite{Braunstein and Caves 1988}} Bell inequality.
A detailed derivation of the various inequalities will not be given, the 
reader is referred to references {\cite{Clauser and Horne 74,Bell 65,Braunstein and Caves 1988}}. 
Here we will consider only strong inequalities, that is inequalities 
where auxiliary assumptions (not based on local realism) are not 
required. In Fig (\ref{fig1}) we depict a very 
idealized setup for general Bell inequality experiment.

Probably the most well known inequality is the Clauser Horne 
strong Bell inequality{\cite{Clauser and Horne 74}} given by
\begin{eqnarray}\label{ch}
|{\bf B}_{ch}|\leq 1
\end{eqnarray}
where
\begin{eqnarray}
B_{ch}={{P_{11}\left(\theta,\phi\right) -P_{11}\left(\theta',\phi\right) +P_{11}\left(\theta,\phi'\right) +
P_{11}\left(\theta',\phi'\right)}\over{ 
P_{1}\left(\theta'\right)+P_{1}\left(\phi\right)}}
\end{eqnarray}
Here $P_{11}$ is the probability that a ``1'' results occurs at each 
analyzer $A,B$ given $\theta,\phi$. Similarly $P_{1}$ is the probability that a ``1'' 
occurs at a detector while having no information about the second. For 
many of the actual experimental considerations an angle factorization 
occurs so that $P_{11}\left(\theta,\phi\right)$ depends only on 
$\theta+\phi$. Also $P_{1}\left(\theta\right)$  and $P_{1}\left(\phi\right)$
are independent of $\theta,\phi$. In this case $B_{ch}$ can be simplified to
\begin{eqnarray}
	B_{ch}={{3 P_{11}\left(\psi\right) -P_{11}\left(3 \psi\right)}\over{ 2 P_{1}}}
\end{eqnarray}
where $\psi=\theta+\phi=-\theta'-\phi'=\theta+\phi'$ and 
$3\psi=\theta'+\phi$.

The second form of the Bell inequality (sometimes referred to as the {\it 
spin} or original Bell inequality) is{\cite{Bell 65}}
\begin{eqnarray}\label{spin}
B_{s}=|E\left(\theta,\phi\right) &-&E\left(\theta',\phi\right) \nonumber \\
&+&E\left(\theta,\phi'\right)+E\left(\theta',\phi'\right)|\leq 2
\end{eqnarray}
where the correlation function $E\left(\theta,\phi\right)$ is given by
\begin{eqnarray}\label{correlation}
E\left(\theta,\phi\right)= 
P_{11}\left(\theta,\phi\right)&+&P_{00}\left(\theta,\phi\right)\nonumber \\
&-&P_{10}\left(\theta,\phi\right)-P_{01}\left(\theta,\phi\right)
\end{eqnarray}
Here as discussed above $P_{11}$ is probability that a ``1'' results occurs at each 
analyzer $A,B$ given $\theta,\phi$. $P_{00}$ is probability that a ``0'' results occurs at each 
analyzer $A,B$, while $P_{10}$ ($P_{01}$) is probability that a 
``1'' (``0'') results occurs at the analyzer $A$ and a ``0'' (``1'') at $B$. 
With the angle factorization given above, the inequality (\ref{spin}) 
can be rewritten as
\begin{eqnarray}\label{spin1}
B_{s}=\left|3 E\left(\psi\right) -E\left(3 \psi\right) \right|\leq 2
\end{eqnarray}

Our final form of Bell inequality to be considered in this article was developed by Braunstein and 
Caves{\cite{Braunstein and Caves 1988}}. This classical information-theoretic 
Bell inequality has the form
\begin{eqnarray}\label{info}
B_{info}\geq 0
\end{eqnarray}
where
\begin{eqnarray}\label{infoB}
B_{info}= 
-H\left(\theta|\phi\right)&+&H\left(\theta|\phi'\right)\nonumber \\&+&
H\left(\phi'|\theta'\right)+H\left(\theta'|\phi\right)
\end{eqnarray}
Here  $H\left(\theta|\phi\right)$ is given by
\begin{eqnarray}
H\left(\theta|\phi\right)= - \sum_{a,b} P(a,b) \log 
\left({{P(a,b)}\over{P(a)}} \right)
\end{eqnarray}
with $\log \left(P(a,b)/P(a) \right)$ being the information gained at $B$ 
given the result at $A$ is known. The conditional information is then 
given by $H\left(\theta|\phi\right)$. The base of the 
logarithm determines the units of the information (base 2 for bits, 
base e for nats). For quantum computing purposes, this inequality 
should prove highly useful as it directly deals with information content. 
Several other Bell inequality do exist such as the CHSH 
inequality{\cite{CHSH 69}}, but these are not considered here due to 
there weaker nature. Auxiliary assumptions are necessary in there derivation which open up 
several loopholes{\cite{Kwiat et al 94,Freyberger et ak 95,Fry et al 96}}.

\section{Correlated States}

From (\ref{correlatedpair}) we need to find the optimal $c_{n}$ 
which gives the largest Bell inequality 
violation. Before determining the $c_{n}$ we need to briefly focus our attention 
on the quadrature phase homodyne measurement.

A quadrature phase-amplitude homodyne measurement $X(\theta)$ at $A$ can achieved by combining
a signal field (say $\hat a$) with a strong local oscillator field (say
$\epsilon$) to form  two new fields given by $\hat c_{\pm}=\left[ \hat a \pm
\epsilon \exp \left( i \theta\right)\right] /\sqrt{2}$. Here $\theta$ is a
phase shift which allows the choice of particular observable to be
measured, for instance choosing $\theta$ as $0$ or $\pi/2$ allows the measurement 
of the conjugate phase variables $X(0)$ and $X(\pi/2)$ respectively. 
The homodyne measurement gives the  photocurrent difference as 
\begin{eqnarray}
I_{d}&=& c_{+}^{\dagger} c_{+}-c_{-}^{\dagger}c_{-} \nonumber \\
&=& \epsilon \left(\hat a e^{-i \theta}+ \hat a^{\dagger} e^{-i
\theta}\right) =\epsilon  X(\theta)
\end{eqnarray}
Performing a measurement on the quadrature phase amplitude $X(\theta)$ at $A$
yields a result $x_{1}(\theta)$ which ranges in size and sign. Similarly a measurement on 
the quadrature phase amplitude $X(\phi)$ at $B$ yields a result $x_{2}(\phi)$. 
For our state given by (\ref{correlatedpair}), the probability of obtaining 
the result $x_{1}(\theta), x_{2}(\phi)$ is simply
\begin{eqnarray}\label{prob}
P_{x_{1}x_{2}}\left(\theta,\phi \right)= 
\left| \langle x_{1}(\theta)| \langle x_{2}(\phi)| \Psi\rangle \right|^{2}
\end{eqnarray}
where 
\begin{eqnarray}
\langle x (\varphi)|n\rangle={1\over{\sqrt{ 2^{n} n! \sqrt{\pi}}}} e^{-i n \varphi} e^{-x_{i}^{2}/2} H_{n}(x_{i}).
\end{eqnarray}
Here $H_{n}(x_{i})$ is the Hermite polynomial and $\varphi$ is the 
phase of the local oscillator. Eqn (\ref{prob}) can be 
explicitly written as
\begin{eqnarray}\label{prob1}
P_{x_{1}x_{2}}\left(\psi\right)&=&\sum_{n,m}  
{{c_{n}c_{m}^{\ast} e^{-i (n-m) \psi}} \over{ 2^{n+m} n!m! \pi}} \nonumber \\ 
&\;&\;\;\;\;\;\;\;\times \prod_{i=1}^{2} e^{-x_{i}^{2}} H_{n}(x_{i})H_{m}(x_{i})
\end{eqnarray}
where $\psi=\theta+\phi$, that is our 
expression depends only on the sum of the individual local angles. 

The probability given by (\ref{prob1}) is for continuous variables. 
The majority of the tests of  quantum mechanics versus local realism require a binary result. 
Hence for a given quadrature measurement $x_i$ we classify the result as ``1'' 
if $x_i\geq 0$ and the mutually exclusive ``0'' if $x_i<0$. Here we 
have set the binning window about $x_i=0$. Where this binning window is located is quite 
arbitrary, but the maximum violation occurs for the value we have selected.

The probability of obtaining both particles in the ``1'' bin is 
\begin{eqnarray}\label{prob11}
P_{11}\left(\psi \right)&=&\int_{0}^{\infty}\int_{0}^{\infty}  dx_{1}  dx_{2}
P_{x_{1}x_{2}}\left(\psi \right) 
\end{eqnarray}
while the probability of obtaining both particles in the ``0'' bin is 
\begin{eqnarray}\label{prob00}
P_{00}\left(\psi \right)&=&\int_{-\infty}^{0}\int_{-\infty}^{0}  dx_{1}  dx_{2}
P_{x_{1}x_{2}}\left(\psi \right)
\end{eqnarray}
The other probabilities such $P_{10}\left(\psi \right), P_{01}\left(\psi \right)$ 
can be calculated in a similar fashion. The probabilities 
formulated above are joint probabilities. Various of the strong Bell 
inequalities also require marginal probabilities of the form
\begin{eqnarray}\label{marprob}
P_{1}\left(\psi \right)&=&\int_{0}^{\infty}\int_{-\infty}^{\infty}  dx_{1}  dx_{2}
P_{x_{1}x_{2}}\left(\psi \right) 
\end{eqnarray}

The above integrals can be easy evaluated using the 
results{\cite{integraltables}}
\begin{eqnarray}
\int_{0}^{\infty}e^{-x^{2}} H_{n}(x) H_{m}(x)&=&{{\pi 
2^{n+m}}\over{n-m}} \left[{\cal F}(n,m) -{\cal F}(m,n) \right] \nonumber\\
&\;&\;\;\;\;\;\;\;\;\;\;\;\;\;\;\;\;(for\;n\neq m )\nonumber\\
\int_{-\infty}^{\infty}e^{-x^{2}} H_{n}(x) H_{m}(x)&=& 2^{n}n! 
\sqrt{\pi} \delta_{n,m} 
\end{eqnarray}
where ${\cal F}(n,m)$ is given by
\begin{eqnarray}
{\cal F}^{-1}(n,m)=\Gamma \left({1\over 2}-{1\over 2} n \right) \Gamma 
\left(-{1\over 2} m \right) 
\end{eqnarray}
with $\Gamma$ being the Gamma function. Performing the integrals for 
(\ref{prob11}) and (\ref{prob00})  we find
\begin{eqnarray}
P_{11}\left(\psi\right)&=&P_{00}\left(\psi\right) \nonumber\\
&=&{1 \over 4} +
\sum_{n>m}{{2^{n+m+1} \pi c_{n}c_{m}^{\ast} } \over{ n!m! (n-m)^{2} }} \nonumber \\
&\;&\;\;\;\times  \left[{\cal F}(n,m) -{\cal F}(m,n) \right]^{2}\cos \left[(n-m)  
\psi\right]
\end{eqnarray}
Similarly Eqn (\ref{marprob}) simplifies to 
\begin{eqnarray}
P_{1}&=&1/2
\end{eqnarray}
which is independent of the sum of the local oscillator angle $\psi$.  It is also simple to calculate the 
correlation function $E(\psi)$
\begin{eqnarray}
E(\psi)&=& \sum_{n>m}{{2^{n+m+3} \pi c_{n}c_{m}^{\ast} } \over{ n!m! 
(n-m)^{2} }}\left[{\cal F}(n,m) -{\cal F}(m,n) \right]^{2} \nonumber \\
&\;&\;\;\;\;\;\;\;\;\;\times \cos \left[(n-m) \psi\right] 
\end{eqnarray}

Given the probabilities $P_{11}$, $P_{00}$, $\ldots$ it is also 
possible to calculate the conditional information $H\left(\theta|\phi\right)$
\begin{eqnarray}
H\left(\theta|\phi\right)=&-&P_{11} \log \left[2 P_{11} \right]-P_{00} \log 
\left[2 P_{00} \right]\nonumber \\
&-&P_{10} \log \left[2 P_{10} \right]-P_{01} \log \left[2 P_{01} \right]
\end{eqnarray}

It is now possible to calculate the Clauser Horne (\ref{ch}) and 
spin (\ref{spin})  and information-theoretic (\ref{info}) Bell inequalities. 
Some insight into the problem can be achieved by a careful examination of 
the term 
\begin{eqnarray}
{{2^{n+m} \pi } \over{ n!m! (n-m)^{2} }} \left[{\cal F}(n,m) -{\cal F}(m,n) \right]^{2}
\end{eqnarray}
which is present in all the joint probability distributions. This 
expression has several interesting features. First, as the difference 
between $n$ and $m$ becomes large, the smaller that the above 
expression contributes to any of the probability distributions. The 
main contribution for the expression comes from the case $m=n\pm 1$. 
Second, when $n-m$ is even, the above expression is zero. Finally,  
as $n$ is large the different between the $n,m=n-1$ and $n+1,m=n$ 
elements for fixed large $n$ vanishes and they reach an asymptotic limit 
which is smaller than the $n=1,m=0$ case. If these higher order $n$ 
terms dominate due to the choice of the $c_{n}$ in the probability 
formula, then the various Bell inequalities can not violated. This 
also has the implication that the mean photon number cannot be high 
if a violation is to occur and hence it is not a macroscopic test of quantum mechanics. 


\section{A simple Case}

To begin our investigations of the Bell inequalities, consider the case 
of we have only two photon pair states that is, 
\begin{eqnarray}\label{idealsimple}
| \Psi \rangle = c_{0} | 0 \rangle | 0 \rangle+ c_{1} | 1 \rangle | 1 \rangle
\end{eqnarray}
where for convenience we choose $c_{n}$ real. We also require 
$c_{0}^{2}+c_{1}^{2}=1$. The joint probability distributions are 
readily calculated and in fact
\begin{eqnarray}
P_{11}\left(\psi\right)=P_{00}\left(\psi\right)&=&{1 \over 4}+ {{c_{0} c_{1} }
\over{ \pi}} \cos \left[\psi\right] \\
P_{10}\left(\psi\right)=P_{01}\left(\psi\right)&=&{1 \over 4}- {{c_{0} c_{1} } \over{ \pi}} \cos \left[\psi\right] \\
\end{eqnarray}
Calculating $B_{ch}$ and $B_{s}$ from  (\ref{ch}) and  (\ref{spin}) we find
\begin{eqnarray}
B_{ch}&=& {1 \over 2}+ {{c_{0} c_{1} } \over{ \pi}}\left\{3 \cos \left[ 
\psi_{0}\right]-\cos \left[3 \psi\right]\right\} \\
B_{s}&=& {{4 c_{0} c_{1} } \over{ \pi}}\left\{3 \cos \left[ 
\psi_{0}\right]-\cos \left[3 \psi\right]\right\} \\
\end{eqnarray}
Optimizing for the angle $\psi$ we find
\begin{eqnarray}
B_{ch}&=& {1 \over 2}+ {{2 \sqrt{2} c_{0} c_{1} } \over{ \pi}} \\
B_{s}&=&  {{ 8 \sqrt{2} c_{0} c_{1} } \over{ \pi}} \\
\end{eqnarray}
that is, $|B_{ch}|\leq 1$ and $|B_{s}|\leq 2$ for all $c_{0}$. No violation of the strong 
Clauser Horne or spin Bell inequality is possible. 

For the information theoretic case we find
\begin{eqnarray}
H\left(\psi\right)=&-&{1 \over 2} \log \left[{1 \over 4}- \lambda^{2} \right]- 
\lambda \log \left[{{1+ 2 \lambda  }\over{1- 2 \lambda }}\right]
\end{eqnarray}
where
\begin{eqnarray}
\lambda  ={{2 c_{0} c_{1} }\over{ \pi}} \cos \left[\psi\right] 
\end{eqnarray}
The informational theoretic Bell inequality is given by $B_{info}=3 
H\left(\psi\right)-H\left(3 \psi\right)\geq 0$. A violation of this 
inequality is possible if $B_{info}< 0$. Unfortunately for all $c_{0} 
c_{1}$ and $\psi$ we have $B_{info}> 0$. 

No violation is possible for any of the Bell inequalities considered 
for the ideal state (\ref{idealsimple}) when the detection scheme 
is based on homodyne quadrature phase measurements. If more correlated photon pairs are present 
can a violation be achieved? The obvious answer is yes, because of the recent work of 
Gilchrist {\it et. al.}{\cite{Gilchrist Deuar and Reid 98}}. The real question is how large this 
violation is?

\section{Numerical Studies}

\begin{figure}
\center{ \epsfig{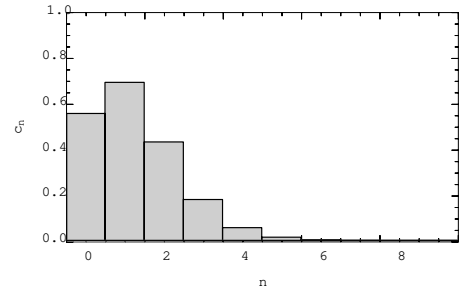}}
\caption{Plot of $c_{n}$ versus $n$}
\label{fig2}
\end{figure}

Considering the expression (\ref{correlatedpair}) for the correlated 
photon pairs, what are the optimal $c_{n}$ coefficients to maximize 
the violation. Because of the results indicated by Gilchrist {\it et. 
al.} and our previous discussion  
we anticipate that the mean photon number per mode must be low to 
obtain a violation. Hence we 
will truncate the number state basis at $10${\cite{stability}} photon 
per mode. Performing a 
numerical optimization over all the $c_{n}$, the optimal set is found 
to maximize the Clauser Horne and spin Bell inequality (Table {\ref{newtable1}}). 
A plot of $c_{n}$ versus $n$ is depicted in Fig (\ref{fig2})

\begin{figure}[h]
\center{ \epsfig{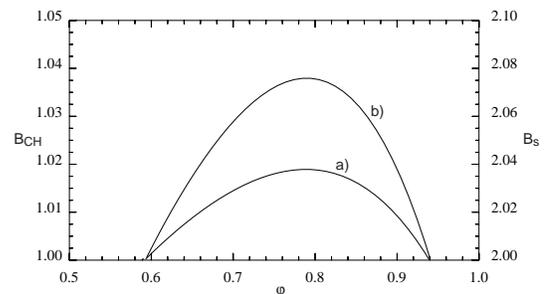}}
\caption{Plot of the Clauser Horne (a) and Spin (b) Bell inequality 
versus $\psi$. A violation occurs for the Clauser Horne Bell 
inequality if $B_{ch}>1$. A violation of the spin Bell inequality occurs 
for $B_{s}>2$.}
\label{fig3}
\end{figure}

It is interesting to now discuss some properties of these optimal $c_{n}$. First the 
general shape of the $c_{n}$ versus $n$ curve shown in (\ref{fig2}) is 
similar to that considered in the circle state by Gilchrist {\it et. 
al.}{\cite{Gilchrist Deuar and Reid 98}}. It is however not exactly 
the same (see (Table {\ref{newtable1}}). Given this optimal parameter set, what is the maximum violation of the 
Bell inequalities we are considering. In Fig (\ref{fig3}) we plot 
both the Clauser Horne and spin Bell inequalities versus $\psi$.

For the Clauser Horne Bell inequality the maximum violation 
corresponds to $B_{ch}=1.019$, while the maximum violation for the 
spin Bell inequality corresponds to  $B_{s}=2.076$. Interesting here 
is that the percentage violation of the spin inequality is approximately 
$3.8\% $ compared with the $1.9\% $ for the Clauser Horne case. This 
significantly increases the potential for an experiment to be 
performed provided such an experiment were not significantly more 
difficult. Also the results for the optimal $c_{n}$ set give a Clauser 
Horne Bell inequality violation that is approximately $20 \%$ greater 
than the circle state results of Gilchrist {\it et. al.}{\cite{Gilchrist Deuar and Reid 98}}.

\begin{figure}
\center{ \epsfig{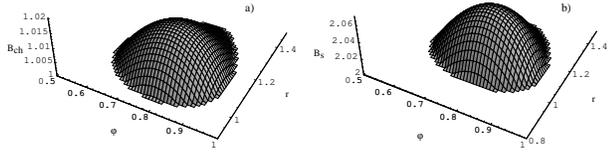}}
\caption{Plot of the Clauser Horne (a) and Spin (b) Bell inequality 
versus $r$ and $\psi$. A violation occurs for the CH Bell 
inequality if $B_{ch}>1$. A violation of the Spin Bell inequality occurs 
for $B_{s}>2$.}
\label{fig4}
\end{figure}

It is interesting to consider whether a greater violation of the Bell 
inequality can be achieved with the state given by (\ref{circlecn}). 
To this end we show the effect of the variation of both $r$ 
and $\psi$ (sum of the local oscillator angles) for both the Clauser 
Horne and spin Bell inequalities in Fig (\ref{fig4}). As can be seen 
the spin Bell inequality can be violated far more significantly than 
the similar Clauser Horne case. In fact, as occurred previously the 
percentage maximum violation in the spin inequality is twice that of 
the Clauser Horne result.

In any of the analysis considered above we have not discussed errors, there 
sources and how they effect the potential violation. We will not 
present any significant details here in this article but refer the reader to 
{\cite{Gilchrist Deuar and Reid 98}} for such a decision.

\begin{figure}
\center{ \epsfig{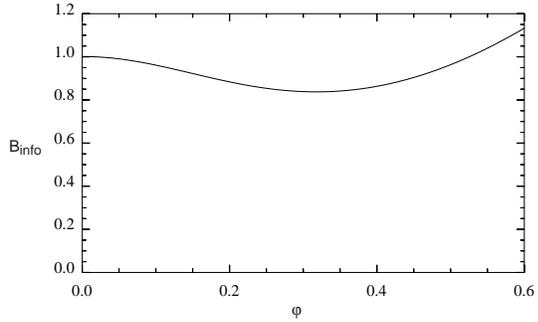}}
\caption{Plot of the information-theoretic Bell inequality 
versus $\psi$. A violation is possible for $B_{info}<0$.}
\label{fig5}
\end{figure}

Our final Bell inequality to be considered is the Braunstein and 
Caves{\cite{Braunstein and Caves 1988}} information-theoretic case. 
In Fig (\ref{fig5}) we plot $B_{info}$ versus $\psi$. No violation of 
the information-theoretic inequality is possible for any $\psi$.

A question here to be addressed is why two of the strong 
inequalities can be violated while this information-theoretic Bell 
inequality is far from  being violated. In the binning process to give 
a binary result for a quadrature measurement, information must be 
discarded. The  information-theoretic inequality is much more 
sensitive to this information loss than the Clauser Horne inequality. 
Also why would we fundamentally expect all three inequalities to be 
violated. A violation of any of the inequalities indicate a 
discrepancy between quantum mechanics and local realism.

\section{Conclusion}
In this article we have place strict bounds on the 
optimal $c_{n}$ coefficients for the state (\ref{correlatedpair}) which maximizes the Clauser Horne and 
Spin Bell inequalities when a homodyne quadrature phase measurements 
is performed. The spin Bell inequality is violated by approximately 
$3.6\%$ while the Clauser Horne inequality is violated by approximately 
$1.9\%$. The violation is small however due to the fact that we are discarding 
information in the binning process. In fact due to the information 
loss in the binning process the information theoretic Bell inequality 
is not violated in any regime. A larger violation cannot be obtained 
using homodyne measurements with the strong inequalities we have considered.

While our optimal $c_{n}$ coefficient give a slightly better 
violation than the pair coherent state, it is difficult to see how 
such a state could be generated. Closely examining the spin Bell 
inequality with the pair coherent state still indicates that a 
greater violation (approximately twice the size) is possible than for 
the other inequalities. This would make the test much more feasible 
provided the pair coherent state could be generated. 
In such a system the mean photon number is small, so this is not 
strictly a macroscopic test of quantum mechanics. It does however have 
a macroscopic nature due the strong local oscillator which means 
large photodetector currents are obtained.

To conclude, quadrature phase homodyne measurement provide a mechanism 
for performing tests of the Bell inequality with highly efficient 
detection. This allows one of the loopholes in current experiments to 
be closed. However, due to the inherent information loss in the 
binning process, the violations are small but should be achievable.

 .

\begin{table}
\caption{The optimal $c_{n}$ parameters to maximize the violation of 
the Clauser Horne and spin Bell inequalities. The  $c_{n}$ values for 
the Gilchrist circle state are also given}
\begin{tabular}{lcc}
n&$c_{n}$&Eqn (\ref{circlecn}) with $r\sim 1.12$\\ \tableline
0& 0.4990& 0.5495 \\
1& 0.6355& 0.6893 \\
2& 0.4760& 0.4323 \\
3& 0.3135& 0.1808 \\
4& 0.1465& 0.0567 \\
5& 0.0235& 0.0142 \\
6& 0.0075& 0.0029 \\
7& 0.0024& 0.0005 \\
\tableline \tableline
$B_{ch}$ & 1.019 & 1.016 \\
$B_{s}$ & 2.076 & 2.064 
\end{tabular}
\label{newtable1}
\end{table}

\end{multicols}


\begin{references}
\bibitem{strong} By a strong test of quantum mechanics versus local 
realism, we mean a test in which no auxiliary assumptions are 
necessary.
\bibitem{stability} Checks were performed to ensure that truncation 
errors were minimal (less than $10^{-6}$).
\end{references}
\end{document}